\documentstyle[preprint,aps,epsfig]{revtex}
\tightenlines
\begin{document}

\title{Exact analytic expression for a subset of fourth virial
coefficients of polydisperse hard sphere mixtures}
\author{Ronald Blaak}
\address{FOM Institute for Atomic and Molecular Physics, Kruislaan 407,\\
1098 SJ Amsterdam, The Netherlands.} 
\date{\today}
\maketitle

\begin{abstract}
{We derive an exact, analytic expression for the fourth virial
coefficient of a system of polydisperse spheres under the constraint
that the smallest sphere has a radius smaller than a given function
of the radii of the three remaining particles.}
\end{abstract}

\section{Introduction}
\label{sec:intro}
It is surprisingly difficult to calculate analytic expressions for
virial coefficients. Even the exact calculation of the second virial
coefficient is in general extremely difficult. If particles have no
other interaction than hard core repulsion, things are slightly
easier. In the isotropic phase, the second virial coefficient $B_2$ of
two arbitrary, convex particles, $A$ and $B$, can be determined by
\cite{Isihara:50}  

\begin{equation}
\label{eq:b2-iso}
B_2(A,B) = \frac{1}{8 \pi} \left( V(A) G(B) + S(A) M(B) + M(A) S(B) +
G(A) V(B) \right) 
\end{equation}
where only geometrical quantities of both particles are used: the
volume $V$, the surface area $S$, and the integrals $M$ and $G$ over
the mean respectively Gaussian curvature. For higher virial
coefficients only approximations of a similar type are known
\cite{Boublik:86}. 

The second and third virial coefficients of identical spheres are
known analytically and are 
obtained after a straightforward calculation. The fourth virial
coefficient, for which the analytic expression due to Boltzmann is also
known \cite{Boltzmann:1899}, is much more difficult. Beyond the fourth
virial coefficient only numerical data are available
\cite{Kratky:77}. More recently, data of the fourth and fifth virial
coefficients of binary mixtures have been published by Saija {\em et
al} \cite{Saija:96-B4,Saija:96-B5,Saija:96-B4er}.

In this paper we will focus on hard spheres with different radii, for
which until now, no analytic results are known beyond the third
virial coefficient. In section \ref{sec:2-3} we show the results for
the second and third virial coefficients. In section \ref{sec:fourth}
we derive the known results for the fourth virial coefficient. The
main problem, the calculation of the complete star diagram, is derived
in section \ref{sec:star} together with an inequality, which has to be
satisfied, in order for the given expression to be valid. In section 
\ref{sec:discussion} we finish by discussing our result and some
concluding remarks.

\section{Second and third virial coefficient}
\label{sec:2-3}

The general expression for the $n$th virial coefficient $B_n$ of a
gas with pairwise, additive interaction $\phi_{ij}$ between particles $i$
and $j$ is given in terms of Mayer $f_{ij}$ functions by
\begin{equation}
\label{eq:Bn}
B_n = \frac{1-n}{n!} \lim_{V \rightarrow \infty} V^{-1} \int \cdots \int V_n 
d{\bf r}_1 \cdots d{\bf r}_n,
\end{equation}
where ${\bf r}_i$ are the spatial coordinates of particle $i$, and $V$
is the volume accessible for the particles. $V_n$ is the sum over all
labeled stars with $n$ points given by
\begin{equation}
\label{eq:sum}
V_n \equiv \sum_{\{S_n \}} \prod_{i>j}^n f_{ij}
\end{equation}
and expressed in the Mayer $f$ functions, which are related to the
interaction potential by
\begin{equation}
\label{eq:Mayer}
f_{ij} \equiv \exp(-\beta \phi_{ij}) - 1.
\end{equation}
For hard particles the Mayer $f$ function reduces to
\begin{equation}
\label{eq:Mayer-hard}
f = \left\{ \begin{array}{cl} 0 & \mbox{if no overlap} \\ -1 &
\mbox{if overlap} 
\end{array} \right.
\end{equation}
because the interaction potential is either zero or infinity in the
case that particles do overlap, respectively do not overlap. 

The second virial coefficient $B_2(A,B)$ for two particles $A$ and $B$,
has only one contributing diagram, and is given by
\begin{equation}
B_2(A,B) = 
\frac{1}{2}\raisebox{-18pt}{\psfig{figure=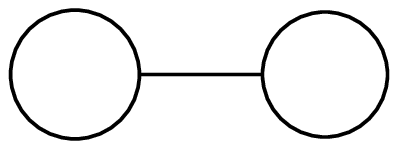,width=.6in}} =
\frac{1}{2}\raisebox{-18pt}{\psfig{figure=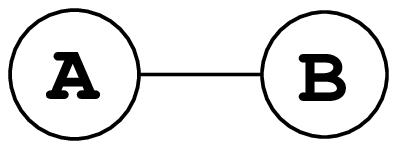,width=.6in}},
\end{equation}
where we have explicitly labeled the diagram. We use diagrams to
represent the different integrals. Integrations are over all possible
positions of the different particles with one particle fixed. Each
solid line represents an 
overlap between the connected particles, and ,hence, some constraint on
the position vectors. The signs, which are determined by the
Mayer functions, are, for convenience, immediately put in front of the
diagrams. The diagram in $B_2$ is the volume from which the second
particle is excluded, in order not to cause an overlap. For 
spheres with radii $A$ and $B$ this is simply the volume of a sphere 
with radius $A+B$, leading to  the well known result
\begin{equation}
B_2(A,B) = \frac{2}{3} \pi (A+B)^3.
\end{equation}
Also the third virial coefficient $B_3(A,B,C)$ of three particles $A$,
$B$ and $C$ has only one contributing diagram

\begin{equation}
B_3(A,B,C) = 
\frac{2}{3!}\raisebox{-18pt}{\psfig{figure=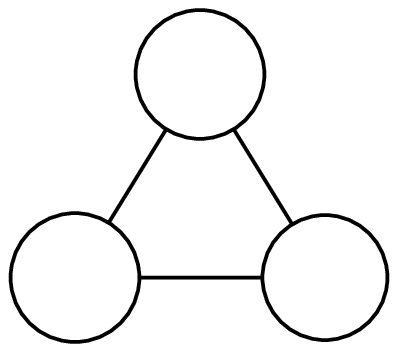,width=.6in}} = 
\frac{1}{3} \raisebox{-18pt}{\psfig{figure=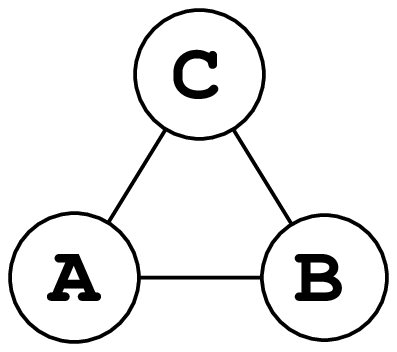,width=.6in}}.
\end{equation}
Only in the case of simple objects, e.g. spheres and discs, are analytic
expressions for the third virial coefficient known.

In order to proceed we will assume that the radii of the three spheres
$A$, $B$ and $C$ satisfy the inequality $A \geq B \geq C$. We
position the biggest particle at the origin and particle $B$ on the
positive $z$-axis. The integral corresponding to the diagram consists
of two contributions. The first contribution comes from the case where
$B$ is completely inside particle $A$, in which case particle $C$ is
only required to overlap with particle $B$, leading automatically to an
overlap with particle $A$. The second contribution comes from the case
where $B$ is overlapping with $A$, but not enclosed. Particle $C$ now
has to overlap with both particles, $A$ and $B$, and hence its center
has to be placed in the overlapping volume of the spheres placed at
$A$ with radius $A+C$ and the sphere placed at $B$ with radius
$B+C$. This volume $Z(z_{AB},A+C,B+C)$ is the sum of two segments of
different spheres and depends on the distance $z_{AB}$ between $A$ and
$B$. The formula $Z(r,R_1,R_2)$ can easily be evaluated if we assume
$R_1 \geq R_2$ 
\begin{equation}
Z(r,R_1,R_2) = \left\{ \begin{array}{l}
\frac{4 \pi}{3} R_2^3 \mbox{\hspace{1cm} $r < R_1 - R_2$}\\
\frac{\pi}{12 r} (R_1 + R_2 - r)^2 (r^2 - 3(R_1 -
R_2)^2 + 2 r (R_1 + R_2)) \\
0 \mbox{\hspace{1.75cm} $r > R_1 + R_2$} \end{array} \right.
\end{equation}
if $r < R_1 - R_2$ the smaller sphere is completely enclosed by the
larger sphere, and, hence, the overlap is equal to the volume of the
smaller sphere. If $r > R_1 + R_2$ the spheres are not overlapping at
all. The diagram for the third virial coefficient is therefore given
by 
\begin{equation}
\raisebox{-18pt}{\psfig{figure=figures/V3ABC.eps,width=.6in}} = 
\int\limits_0^{A+B} d z_{AB} 4 \pi z_{AB}^2 Z(z_{AB},A+C,B+C)
\end{equation}
which results for the third virial coefficient $B_3(A,B,C)$ in
\begin{eqnarray}
B_3(A,B,C) = \frac{16 \pi^2}{27} ( A^3 B^3 + B^3 C^3 + A^3 C^3
+  \nonumber \\ 
  3 A B C [ A + B + C ] [A B + B C + C A] ) .
\end{eqnarray}
Although we have assumed that $A \geq B \geq C$, the resulting formula
is fully symmetric in the radii.

\section{Fourth virial coefficient}
\label{sec:fourth}
In order to obtain the fourth virial coefficient $B_4$ for particles $A$,
$B$, $C$ and $D$, we need to evaluate three different diagrams,
although this could be reduced to only two modified star diagrams (see
ref. \cite{Ree:64})
\begin{equation}
\label{eq:B4}
B_4(A,B,C,D) = -\frac{3}{4!} \left( 
3 \raisebox{-18pt}{\psfig{figure=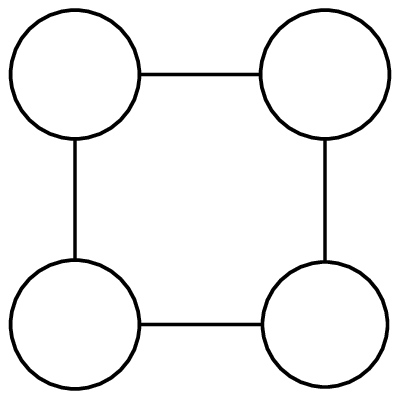,width=.6in}} -
6 \raisebox{-18pt}{\psfig{figure=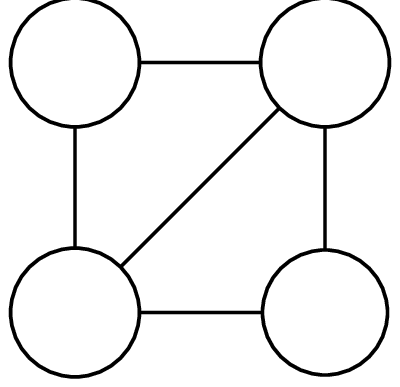,width=.6in}} +
  \raisebox{-18pt}{\psfig{figure=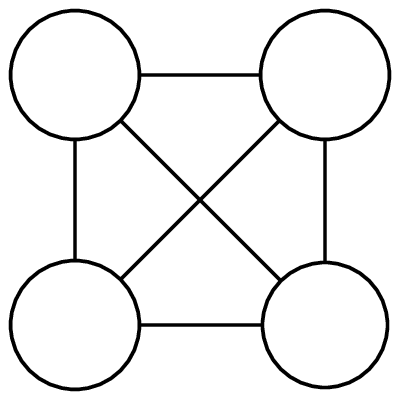,width=.6in}}
\right).
\end{equation}
The fourth virial coefficient is here expressed in unlabeled diagrams.
However, labeling now becomes important because we want to evaluate
those diagrams for the case of non-identical particles. This leads to
the following expressions
\begin{equation} \label{eq:V4} \begin{array}{lll}
3 & \raisebox{-18pt}{\psfig{figure=figures/V40.eps,width=.6in}} = &
    \raisebox{-18pt}{\psfig{figure=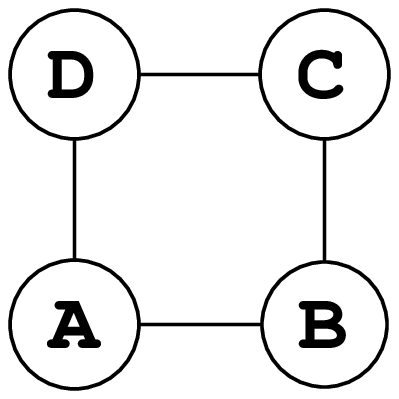,width=.6in}} +
    \raisebox{-18pt}{\psfig{figure=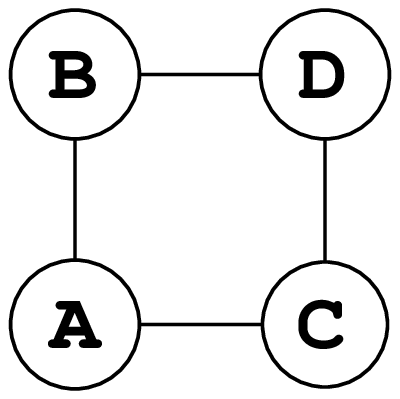,width=.6in}} +
    \raisebox{-18pt}{\psfig{figure=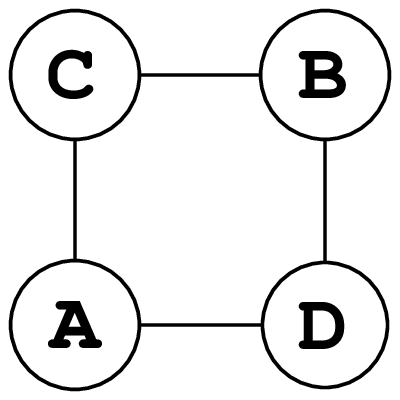,width=.6in}}
\\
6 & \raisebox{-18pt}{\psfig{figure=figures/V41.eps,width=.6in}} = &
    \raisebox{-18pt}{\psfig{figure=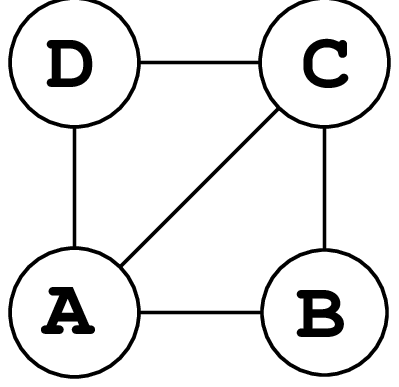,width=.6in}} +
    \raisebox{-18pt}{\psfig{figure=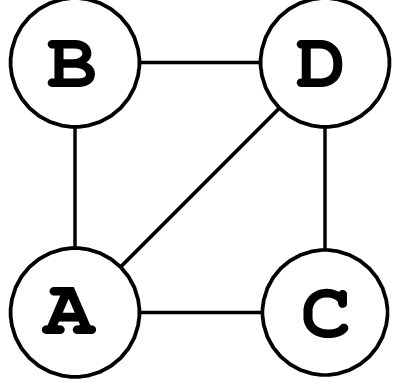,width=.6in}} +
    \raisebox{-18pt}{\psfig{figure=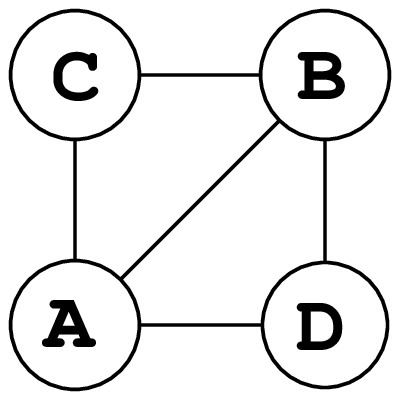,width=.6in}} +
    \raisebox{-18pt}{\psfig{figure=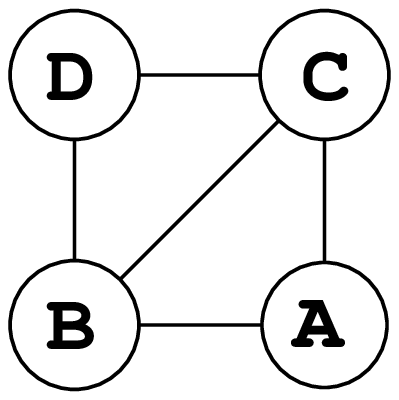,width=.6in}} +
    \raisebox{-18pt}{\psfig{figure=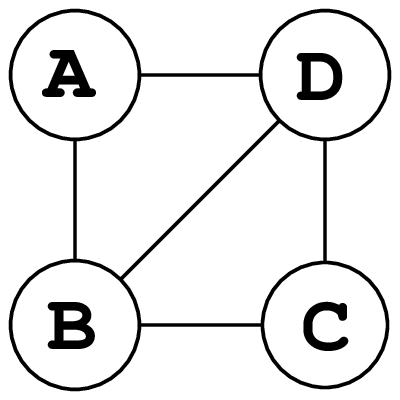,width=.6in}} +
    \raisebox{-18pt}{\psfig{figure=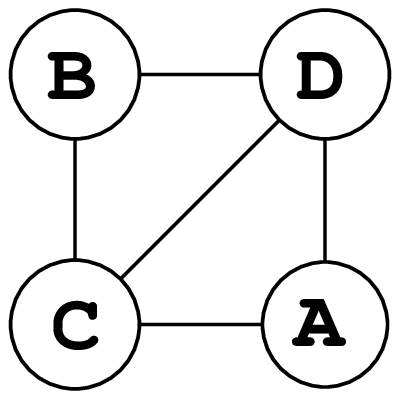,width=.6in}}
\\
  & \raisebox{-18pt}{\psfig{figure=figures/V42.eps,width=.6in}} = &
    \raisebox{-18pt}{\psfig{figure=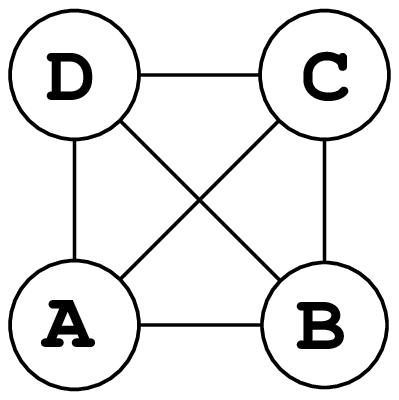,width=.6in}} 
\end{array} \end{equation}

In order to proceed, we need to assume that the radii of the spheres
satisfy $A \geq B \geq C \geq D$. The labeled diagrams in the first
line of (\ref{eq:V4}) are given by expressions of the form
\begin{equation}
\label{eq:B40}
\raisebox{-18pt}{\psfig{figure=figures/V40ABCD.eps,width=.6in}} = 
\int\limits_0^{A+C+2 D} d z_{AC} 4 \pi z_{AC}^2 Z(z_{AC},A+B,C+B)
Z(z_{AC},A+D,C+D).
\end{equation}
The limits of integration are a consequence of the differences in
size and in this case determined by the fact that $B>D$. The diagrams
in the second 
line of (\ref{eq:V4}) are similar, 
and only differ in the integration limits because of the extra
overlap. They are of the form
\begin{equation}
\label{eq:B41}
\raisebox{-18pt}{\psfig{figure=figures/V41ABCD.eps,width=.6in}} = 
 \int\limits_0^{A+C} d z_{AC} 4 \pi z_{AC}^2 Z(z_{AC},A+B,C+B)
Z(z_{AC},A+D,C+D).
\end{equation}
The other diagrams are obtained by permutation symmetry.
The resulting integrals of (\ref{eq:B40}) and (\ref{eq:B41}) can
easily be evaluated, but lead to lengthy expressions. The summations
over the different labelings do not lead to symmetric expressions in
terms of the radii. If we however take the combination of the diagrams
according to the definition of the fourth virial coefficient
(\ref{eq:B4}), we obtain 

\begin{eqnarray}
\label{eq:simple}
3 \raisebox{-18pt}{\psfig{figure=figures/V40.eps,width=.6in}} -
6 \raisebox{-18pt}{\psfig{figure=figures/V41.eps,width=.6in}} = 
- \frac{64 \pi^3}{9} ( 
A^3 B^3 C^3 + A^3 B^3 D^3 + A^3 C^3 D^3 + B^3 C^3 D^3 + 
 \nonumber \\ 
   3 A B C D [A B + A C + B C + A D + B D + C D][A B C + A B D
+ A C D + B C D] ),
\end{eqnarray}
which is symmetric in the different radii of the spheres.

\section{The complete star diagram}
\label{sec:star}
The problem of calculating the fourth virial coefficient lies in the
remaining diagram, the complete star diagram in which all pairs of the
four particles overlap. An analytic expression for this diagram in
general is not known. What is known is the special case of four
identical radii, which is due to Boltzmann
\cite{Boltzmann:1899,Happel:6}. 
\begin{equation}
\raisebox{-18pt}{\psfig{figure=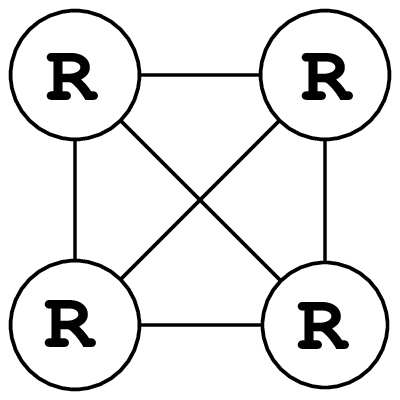,width=.6in}} = 
\frac{256 \pi^2 R^9}{945} \left( 3419 \pi - 438 \sqrt{2} - 8262
\arccos(1/\sqrt{3}) \right),
\end{equation}
where we used $R$ to denote the radius of the particles.

Some limiting cases for different radii are easily obtained. In the
limit that the radius of the smallest sphere goes to 
zero the three remaining particles have to overlap with a point
\begin{equation}
\label{eq:lim1}
\lim_{D \to 0} 
\raisebox{-18pt}{\psfig{figure=figures/V42ABCD.eps,width=.6in}} = 
\left( \frac{4 \pi}{3}\right)^3 A^3 B^3 C^3.
\end{equation}
In the limit where the radius of the biggest sphere goes to infinity,
the other three particles give rise to the diagram related to $B_3$
\begin{equation}
\label{eq:lim2}
\lim_{A \to \infty} 
\raisebox{-18pt}{\psfig{figure=figures/V42ABCD.eps,width=.6in}}
\approx \frac{4 \pi}{3} A^3 
\raisebox{-18pt}{\psfig{figure=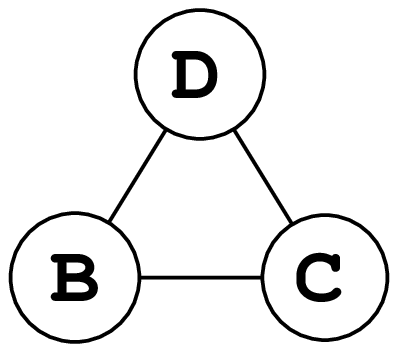,width=.6in}}.
\end{equation}
In the limit where both $A$ and $B$ go to infinity
\begin{equation}
\label{eq:lim3}
\lim_{A,B \to \infty} 
\raisebox{-18pt}{\psfig{figure=figures/V42ABCD.eps,width=.6in}}
\approx \lim_{A,B \to \infty} 
 \raisebox{-18pt}{\psfig{figure=figures/V3ABC.eps,width=.6in}}
\frac{4 \pi}{3} (C+D)^3 \approx \frac{64 \pi^3}{27} A^3 B^3 (C+D)^3.
\end{equation}
In order to explore the star diagram for non-limiting cases we must
look in more detail to the diagrams and introduce the $\tilde f$-bonds
as defined by Ree and Hoover \cite{Ree:64} which we denote by a dashed
line. Such a connection refers to non-overlapping particles, solid
connections to overlapping particles and no connection allows for
both, overlap and no overlap.  

We will now consider diagrams in which
the particles $A$, $B$ and $C$ all overlap with the smallest particle
$D$. The diagram without other constraints can be written as the
following summation of diagrams:
\begin{eqnarray}
\raisebox{-18pt}{\psfig{figure=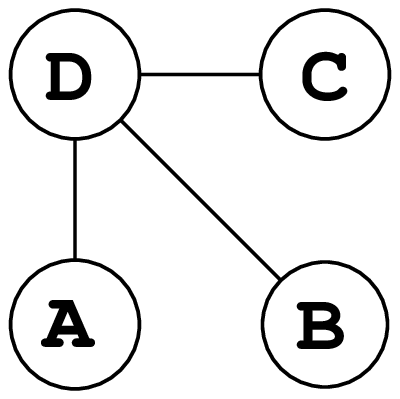,width=.6in}} = &
\raisebox{-18pt}{\psfig{figure=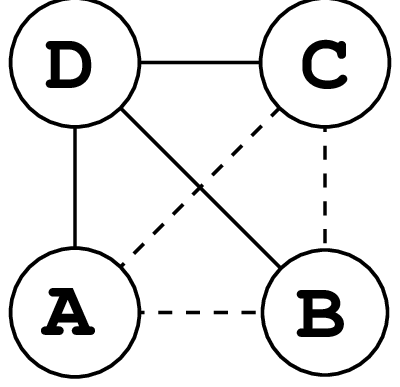,width=.6in}} +
\raisebox{-18pt}{\psfig{figure=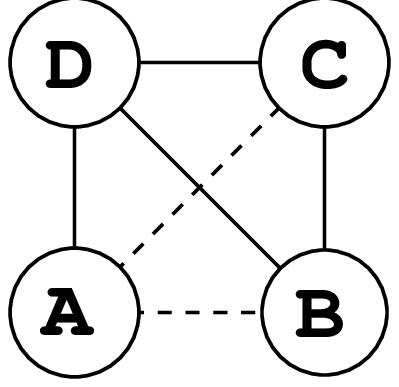,width=.6in}} +
\raisebox{-18pt}{\psfig{figure=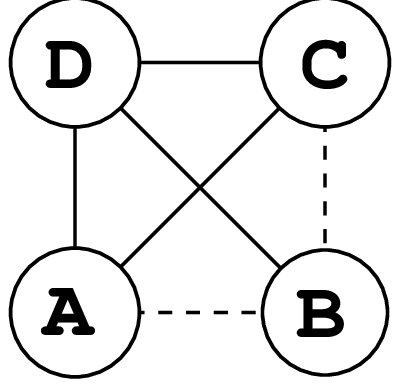,width=.6in}} +
\raisebox{-18pt}{\psfig{figure=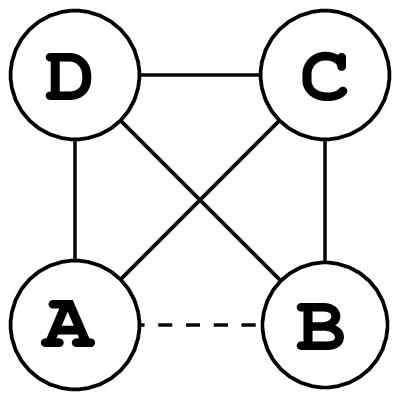,width=.6in}} + 
\nonumber \\ &
\raisebox{-18pt}{\psfig{figure=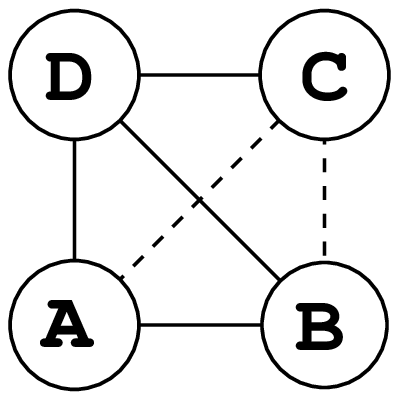,width=.6in}} +
\raisebox{-18pt}{\psfig{figure=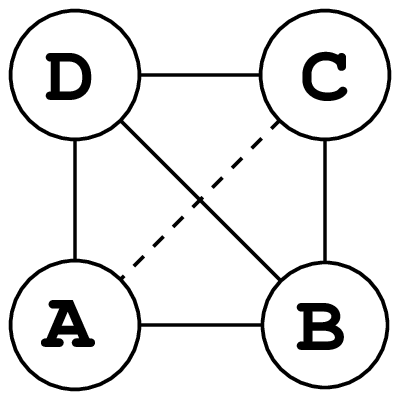,width=.6in}} +
\raisebox{-18pt}{\psfig{figure=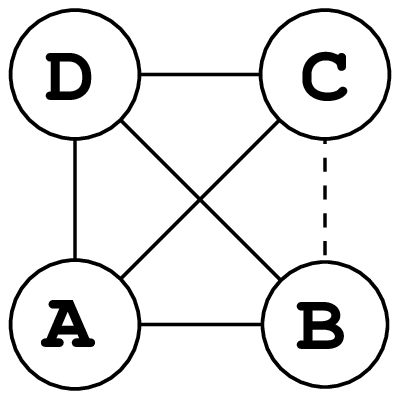,width=.6in}} +
\raisebox{-18pt}{\psfig{figure=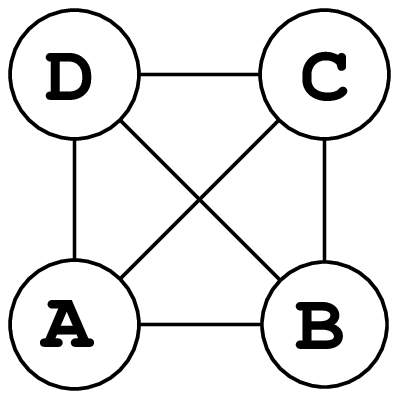,width=.6in}},
\end{eqnarray}
where we have denoted at the right hand side explicitly whether pairs
of particles overlap or not. The diagram at the left hand side can,
however, be evaluated immediately 

\begin{equation}
\raisebox{-18pt}{\psfig{figure=figures/SA.eps,width=.6in}} = 
\raisebox{-18pt}{\psfig{figure=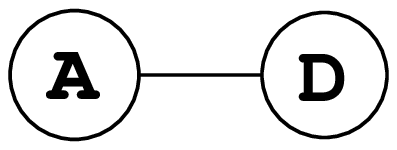,width=.6in}} \times
\raisebox{-18pt}{\psfig{figure=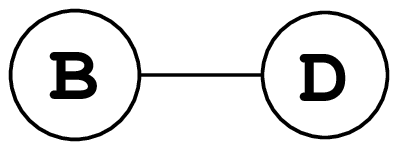,width=.6in}} \times
\raisebox{-18pt}{\psfig{figure=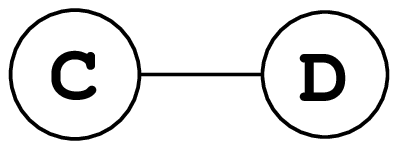,width=.6in}}.
\end{equation}
The same is true for combinations of the form:
\begin{equation}
\raisebox{-18pt}{\psfig{figure=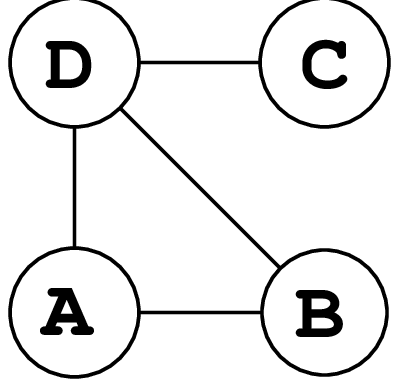,width=.6in}} = 
\raisebox{-18pt}{\psfig{figure=figures/S4.eps,width=.6in}} +
\raisebox{-18pt}{\psfig{figure=figures/S5.eps,width=.6in}} + 
\raisebox{-18pt}{\psfig{figure=figures/S6.eps,width=.6in}} + 
\raisebox{-18pt}{\psfig{figure=figures/S7.eps,width=.6in}} = 
\raisebox{-18pt}{\psfig{figure=figures/V2CD.eps,width=.6in}} \times
\raisebox{-18pt}{\psfig{figure=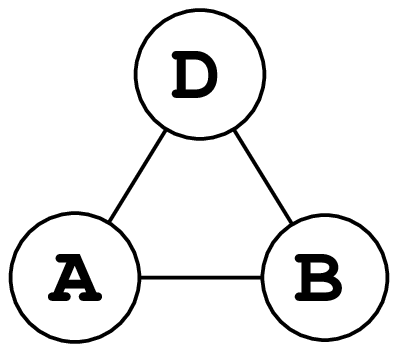,width=.6in}}.
\end{equation}
As well, some combinations have already been evaluated before
(\ref{eq:B41}), e.g. 
\begin{equation}
\raisebox{-18pt}{\psfig{figure=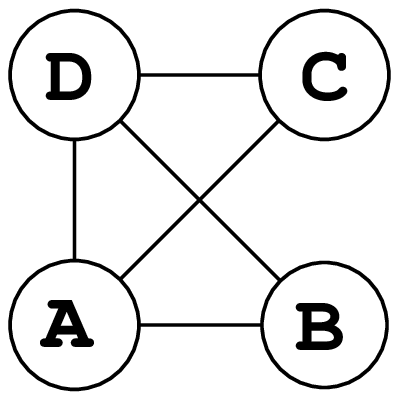,width=.6in}} = 
\raisebox{-18pt}{\psfig{figure=figures/S6.eps,width=.6in}} +
\raisebox{-18pt}{\psfig{figure=figures/S7.eps,width=.6in}} =
\raisebox{-18pt}{\psfig{figure=figures/V41ACDB.eps,width=.6in}}.
\end{equation}
This leads to the following expression for the complete star diagram
\begin{eqnarray}
\label{eq:star}
\raisebox{-18pt}{\psfig{figure=figures/S7.eps,width=.6in}} = &
\raisebox{-18pt}{\psfig{figure=figures/SA.eps,width=.6in}} -
\raisebox{-18pt}{\psfig{figure=figures/S4567.eps,width=.6in}} -
\raisebox{-18pt}{\psfig{figure=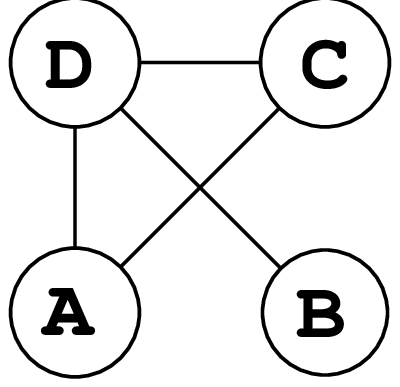,width=.6in}} -
\raisebox{-18pt}{\psfig{figure=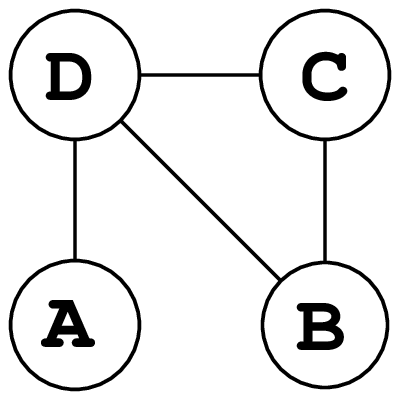,width=.6in}} +
\nonumber \\ &
\raisebox{-18pt}{\psfig{figure=figures/S67.eps,width=.6in}} +
\raisebox{-18pt}{\psfig{figure=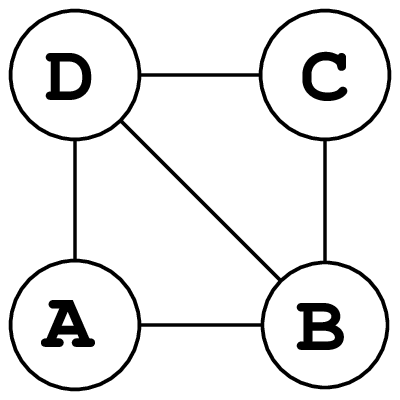,width=.6in}} +
\raisebox{-18pt}{\psfig{figure=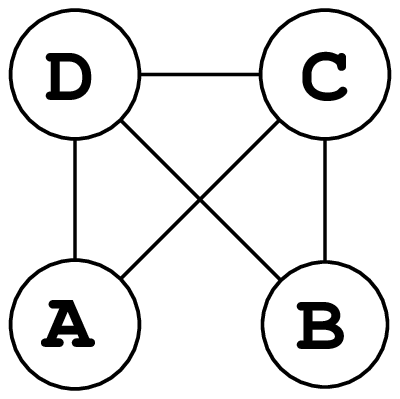,width=.6in}} -
\raisebox{-18pt}{\psfig{figure=figures/S0.eps,width=.6in}}.
\end{eqnarray}
For all diagrams on the right hand side we have an analytic expression
with only one exception - the last diagram. We demonstrate below that
this last diagram is identical to zero if the radii of the spheres
satisfy a simple inequality. 

\begin{center}
\begin{figure}[ht]
\epsfig{figure=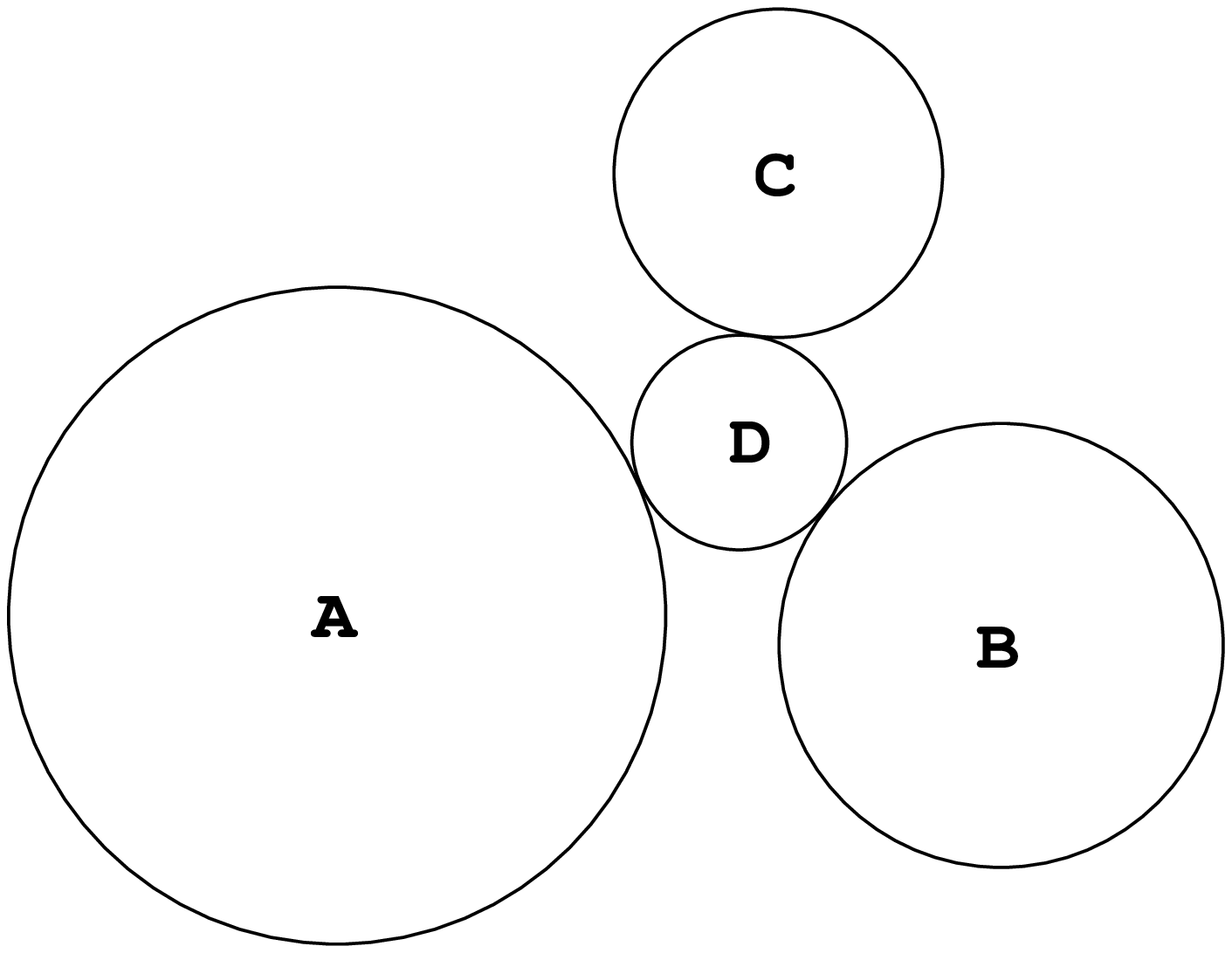,width=6cm}
\vspace{3cm}
\caption[a]{Blaak\label{fig:Apollonian}}
\end{figure}
\end{center}

In order for this diagram to be non-zero the spheres with radius $A$, $B$
and $C$ should all overlap with the smallest sphere with radius $D$,
but have no overlaps with each other. If the spheres $A$, $B$ and $C$
are mutually non-overlapping as shown in figure \ref{fig:Apollonian},
there is a smallest sphere $D$ that can be constructed that will touch
with all three spheres. The centers of mass must all lie in
the same plane, and therefore this problem reduces to the one of
touching circles, known as Apollonian circles since the construction
of this fourth circle was first solved by the Greek mathematician
Apollonius of Perga (200 BC) \cite{Coxeter:68}. If the radius $D$ were
smaller, at least one of the spheres $A$, $B$ or $C$ would have no
overlap with this circle.  

\begin{center}
\begin{figure}[ht]
\epsfig{figure=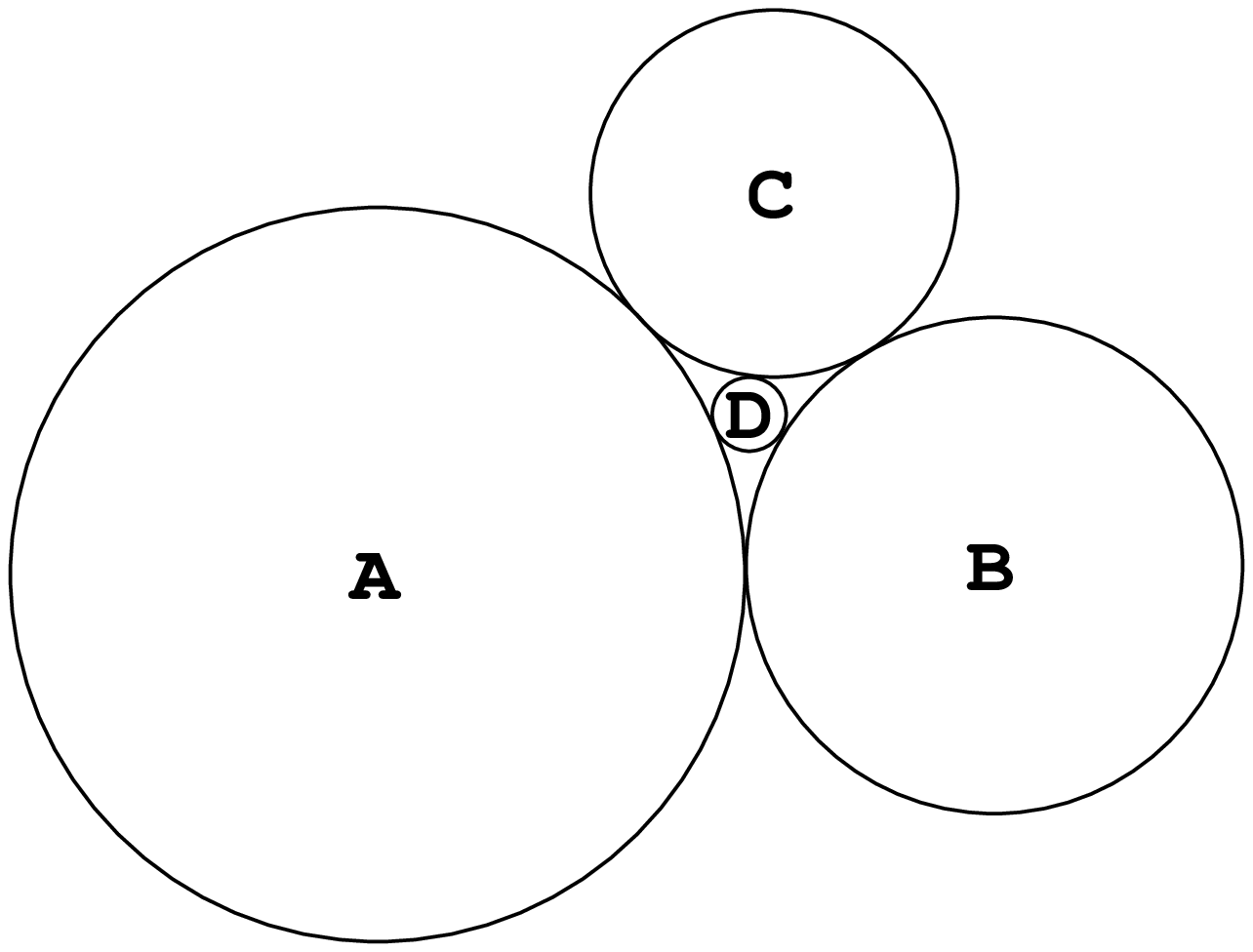,width=6cm}
\vspace{3cm}
\caption[a]{Blaak\label{fig:Apollonian_Dec}}
\end{figure}
\end{center}

If this were the case for all possible configurations of non-overlapping
spheres $A$, $B$ and $C$ there would be no contribution to the last
diagram of (\ref{eq:star}), and hence we would know the fourth virial
coefficient. The smallest
radius $D$ which can contribute is found in the case that all four
spheres touch as depicted in figure \ref{fig:Apollonian_Dec}. For this
special case of the Apollonian problem, a simple relation between the
radii of the circles was derived by the French philosopher and
mathematician Descartes \cite{Coxeter:68}, known as the Descartes circle
theorem: 
\begin{equation}
\label{eq:Decartes}
\left(\frac{1}{A}+\frac{1}{B}+\frac{1}{C}+\frac{1}{D}\right)^2 = 
2 \left(\frac{1}{A^2}+\frac{1}{B^2}+\frac{1}{C^2}+\frac{1}{D^2}\right).
\end{equation}
Solving this equation gives the upper limit for radius $D$ 
\begin{equation}
\label{eq:maxD}
D \leq \frac{A B C}{A B + A C + B C + 2 \sqrt{A B C (A + B + C)}},
\end{equation}
which will lead to a zero-valued diagram, and hence enables us to give
an analytic expression for the complete star diagram. Therefore we
obtain an analytic expression for the fourth virial coefficient of
polydisperse spheres under the constraint (\ref{eq:maxD})
\begin{equation} \begin{array}{l}
B_4(A,B,C,D) =  \left( \frac{16 \pi^3}{27} \right) \times (
A^3 B^3 C^3 + A^3 B^3 D^3 + A^3 C^3 D^3 + B^3 C^3 D^3 + \\
~~~   3 A B C D [ A B + A C + B C + A D + B D + C D][A B C + A B D
 + A C D + B C D]) - \\
~~~\frac{16 \pi^3}{3}  D^3 A^2 B^2 C^2 +
\frac{ 8 \pi^3}{3}  D^4 A B C (A B + A C + B C) -\\
~~~\frac{ 8 \pi^3}{15} D^5 (A^2 B^2 - 2 A^2 B C - 2 A B^2 C + A^2 C^2 - 
2 A B C^2 + B^2 C^2) - \\
~~~\frac{ 8 \pi^3}{5}  D^6 (A + B) (A + C) (B + C) -
\frac{32 \pi^3}{35} D^7 (A^2 + 3 A B + B^2 + 3 A C + 3 B C + C^2) - \\
~~~\frac{ 8 \pi^3}{7}  D^8 (A + B + C) - 
\frac{ 8 \pi^3}{21} D^9 
\end{array} \end{equation}
The first part is symmetric in the four different radii and
proportional to (\ref{eq:simple}), the last part, however, is only
symmetric in the radii $A$, $B$ and $C$. Note that the limiting cases 
(\ref{eq:lim1}), (\ref{eq:lim2}) and (\ref{eq:lim3}) are in agreement
with this result. 

In the case that $A=B=C=1$, and, since according to (\ref{eq:maxD}) $D
\leq (2\sqrt{3} - 3)/3 \approx 0.1547$, this formula reduces to 
\begin{eqnarray}
B_4(1,1,1,D) = & \left( \frac{8 \pi^3}{945} \right) \left( 70 + 630 D +
2520 D^2 + 1470 D^3 + 945 D^4  + \right. \nonumber \\ & \left. 
 189 D^5 - 1512 D^6 - 1296 D^7  - 405 D^8  - 45 D^9 \right).
\end{eqnarray}
This result can be compared with the numerical data for the binary
mixture \cite{Saija:96-B4er}. Only the first three numbers in the
first column satisfy the inequality (\ref{eq:maxD})
\begin{equation}
\begin{array}{llll}
D_{1112}({\cal R}=\frac{1}{20}) & = \frac{11058144323491
\pi^3}{6193152000000000} & = 0.0553630659 & ~~~[0.05539(2)]\\
D_{1112}({\cal R}=\frac{2}{20}) & = \frac{31952948861
\pi^3}{12096000000000}   & = 0.0819065785 & ~~~[0.08189(2)] \\
D_{1112}({\cal R}=\frac{3}{20}) & = \frac{23207369313073
\pi^3}{6193152000000000} & = 0.1161886732 & ~~~[0.11617(3)]
\end{array}
\end{equation}
The numerical values obtained by Saija {\em et al.}
\cite{Saija:96-B4er} are given in square brackets.

\section{Discussion}
\label{sec:discussion}
To our knowledge this is the first time that an exact and analytic
solution of the fourth virial coefficient of a hard sphere mixture is
found. One should note, however, that in order for the expression to
be valid, the radius of the smallest sphere is at most 0.1547 times
that of the largest sphere. Our result agrees perfectly with the
relevant, numerical results of Saija {\em et al} \cite{Saija:96-B4er}.
However, in the description of a polydisperse mixture of spheres, the
most interesting case is that of a few large and many small
particles. Unfortunately, this also means that the inequality
(\ref{eq:maxD}) is, in general, not satisfied. The inequality
requires that the smallest radius is strictly smaller than the other
three, which, for instance in a binary mixture, allows only one of the
three mixed virial coefficients to be evaluated.

The method described in this article can almost certainly be used to
evaluate the fourth virial coefficient of other particles,
e.g. discs, and can possibly be used to simplify calculations of
higher virial coefficients of asymmetric mixtures.

\section*{Acknowledgements}
We thank Bela Mulder, Jos\'e Cuesta and James Polson for a critical
reading of the manuscript. The work of the FOM Institute is part of
the research program of FOM and is made possible by financial support
from the  Netherlands Organization for Scientific Research (NWO).

\newpage
\begin{center}
{\small \bf FIGURE CAPTIONS}
\end{center}

\begin{enumerate}
\item The Apollonian problem of constructing a circle $D$ which is
tangent to three arbitrary spheres $A$, $B$ and $C$.
\item A special case of the Apollonian problem in which all spheres
are tangent to each other, and for which case a simple relation
(\ref{eq:Decartes}) between the radii was derived by Descartes.
\end{enumerate}

\end{document}